\documentclass{ws06-procs}

\begin{document}

\title{Implementation of the trigger algorithm for the NEMO project}

\author{d'Angelo G.$^1$, Riccio G.$^{1,2}$, Brescia M.$^{3,2}$}

\address{1 - Department of Physical Sciences, University of Napoli Federico II,
via Cinthia 6, 80126 - Napoli, Italy, E-mail: dangelo@na.infn.it\\
2 - INFN - Napoli unit, via Cinthia 6,  80126 - Napoli\\
3 - INAF - Astronomical Observatory of Capodimonte, via Moiariello 16. 80131 - Napoli, Italy}  

\maketitle

\abstracts{
We describe the implementation of trigger algorithm specifically tailored on the 
characteristics of the neutrino telescope NEMO. Extensive testing against realistic
simulations shows that, by making use of the uncorrelated nature of the noise produced
mainly by the decay of $^{40}K~\beta$--decay, this trigger is capable to discriminate 
among different types of muonic events.}

\section{Introduction}

\begin{figure}[ht]\label{fig:griglia-torri}
\centerline{\epsfxsize=5in\epsfbox{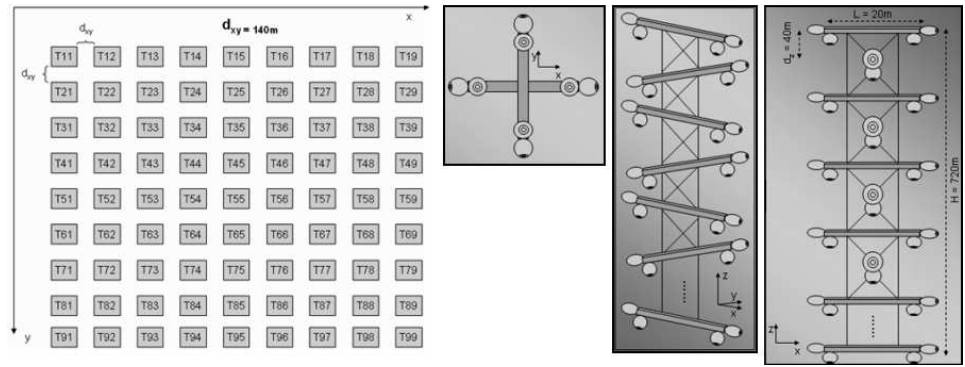}}   
\caption{NEMO telescope. Left panel: general layout. 
Right panel: structure of an individual tower (top and lateral views). }
\end{figure}

The NEMO (NEutrino Mediterranean Observatory) project aims at the deployment of a $1 \ km^3$ 
underwater telescope specifically designed to investigate the properties of cosmic neutrinos 
having energies in the range $10^{19}$ and $10^{22} \ eV $~\cite{sig,ber}. 
The neutrinic component, unlike other components of the cosmic rays, preserves the direction 
of propagation over long distances and therefore it allows to identify the astrophysical 
counterparts responsible for its emission \cite{lip}. 
The telescope will be deployed at about $3500 \ m$ of depth in the Central Mediterranean Sea, 
south of Capo Passero in Sicily. 
It will consist in an orderly grate of 5832 Photo Multiplier Tubes (PMT)\cite{pmt}, organized 
in a grid of $9 \times 9$ towers (Fig. \ref{fig:griglia-torri} a), each composed by 18 floors 
carrying 4 PMTs (Fig. \ref{fig:griglia-torri} b).
The PMTs used for the KM3 (Hamamatsu R7081SEL), will reveal the photons 
produced by Cherenkov effect\cite{fra,gan} and have a spectral response comprised  
between $350$ and $550 \ nm$.

The main difficulties encountered in the detection of the muonic events 
are related to the need to discriminate  the true events from the strong 
background noise which, in the case of NEMO, is mainly due to the Cherenkov
radiation by electrons released by the $\beta$--decay of the potassium--40:

\[^{40}K\rightarrow^{40}Ca + e^- + \overline{\nu}_{e}\]

We implemented a trigger software (which could also be implemented in firmware) 
which can reveal the presence of a muonic event also on a semi real time basis, thus allowing
to issue a target of opportunity alert to ground based and space borne instruments for follow-up
observations. In what follows we shall shortly outline the main characteristics 
of such trigger and its performances as derived from the analysis of simulated data.

\section{Trigger}

The main idea behind the trigger is based on the statistic evidence that 
the PMTs event rate (number of times that a PMT turns on per unit of time) 
is larger for photo multiplier tubes turned on by muonic signals than for 
PMT's turned on by $^{40}K$ events.

From the data acquisition point of view, the NEMO data can be considered as a data stream where 
PMTs are sampled at regular intervals and the signal is set to 1 for PMTs which are turned on and
to 0 for PMTs which are turned off. Each sampling epoch defines what we call a {\it datacube}.
obtained by integrating the data stream over an arbitrarily chosen number of epochs. 
Therefore the implemented trigger makes use of three parameters which can be set by the operator
accordingly to the needs of a specific experiment. 
Namely:

\begin{enumerate}
	\item \textit{sampling time:} time interval between two subsequent readings of the 
	PMTs status (minimum value, $5 \ ns$);
	\item \textit{number of datacubes checked or N:} parameter that characterizes the length of the data stream
	which needs to be checked in order to assess the significance of an event;
	\item \textit{threshold:} event rate threshold for the $^{40}K$ PMTs (minimum value, 1).
\end{enumerate}

In each datacube the trigger eliminates PMTs with rates smaller than the assumed threshold 
but before eliminating it (as due to $^{40}K$) the algorithm checks whether in the N datacubes 
the PMT does not turn on again. 

\section{Test on simulated data}
 
In order to test the trigger, we used realistic (events $+$ noise) 
simulations produced using the GEANT package\cite{gea} as
described in ~\refcite{rdg}. 
The analysis of 126 events using 24 different combinations of the three above mentioned 
parameters\cite{rdg} lead to identify, as best compromise, the following instrumental setup:

\begin{itemize}
	\item \textit{sampling time} = $5 \ ns$;
	\item \textit{number of datacubes checked} = $5$;
	\item \textit{threshold} = $1$.
\end{itemize}

\begin{figure}[ht]  
\centerline{\epsfxsize=5in\epsfbox{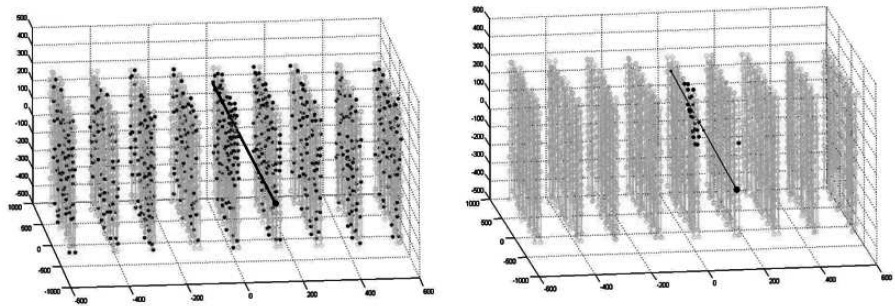}}   
\caption{NEMO telescope. Left panel: the PMTs status integrated over the whole data stream. Right panel: the 
PMTs which survive the pruning performed by the trigger.}\label{fig:datacubes}
\end{figure}

The results are summarized in Fig. \ref{fig:datacubes} which gives: (left panel) the raw status 
of the NEMO telescope, as integrated over the whole simulated event and (right panel) the location 
of the PMTs which are selected by the algorithm as triggered by real events. 
The darker dots represent the active PMTs (due to both muonic and/or $^{40}K$ event), while the lighter ones 
represent the inactive PMTs. The solid line gives instead the event trajectory as derived from the 
simulation. Obviously this knowledge is not in any way used by the trigger and is included only as
a reference for visual inspection. It needs to be stressed that, as it will be further explained in \refcite{rdg},
the efficiency of algorithm is very high, even for short track and or low energy signals.

\section{Future development}

In December 2006, the NEMO project foresees the deployement at a depth of ca. $2000 \ m$ of a tower prototype 
consisting of 4 floors (NEMO phase 1). In this phase all devices and algorithms for detection and reconstruction, 
will be tested.  
We therefore plan to test the trigger on the single tower data in order to assess whether 
it can be implemented in the experiment pipeline. A further development aimed at better reconstructing 
the muon track will require the optimization of the algorithm with respect to the spatial correlation 
of the muonic signal. 


\end{document}